**Aqueous Ion Trapping and Transport in Graphene-embedded 18-crown-6 Ether Pores**

Alex Smolyanitsky*, Eugene Paulechka, Kenneth Kroenlein

Applied Chemicals and Materials Division,
National Institute of Standards and Technology
Boulder, CO 80305

*To whom correspondence should be addressed: alex.smolyanitsky@nist.gov

**Abstract**

Using extensive room-temperature molecular dynamics simulations, we investigate selective aqueous cation trapping and permeation in graphene-embedded 18-crown-6 ether pores. We show that in the presence of suspended water-immersed crown-porous graphene, $K^+$ ions rapidly organize and trap stably within the pores, in contrast with $Na^+$ ions. As a result, significant qualitative differences in permeation between ionic species arise. The trapped ion occupancy and permeation behaviors are shown to be highly voltage-tunable. Interestingly, we demonstrate the possibility of performing conceptually straightforward ion-based logical operations resulting from controllable membrane charging by the trapped ions. In addition, we show that ionic transistors based on crown-porous graphene are possible, suggesting utility in cascaded ion-based logic circuitry. Our results indicate that in addition to numerous possible applications of graphene-embedded crown ether nanopores, including deionization, ion sensing/sieving, and energy storage, simple ion-based logical elements may prove promising as building blocks for reliable nanofluidic computational devices.

Crown ethers[1] are a family of electrically neutral cyclic ethylene oxide molecules, which exhibit marked selective affinity for metal cations. A representative example of a crown ether is shown in Fig. 1 (a). The underlying nearly circular van der Waals coordination allows for a rich variety of "host-guest" binding preferences, depending on the crown size and composition. Unsurprisingly, after the initial discovery of crown ethers more than half a century ago, numerous intriguing applications were proposed, ranging from nanoscale self-assembly [2] to cation sensing and separation.[3-5]

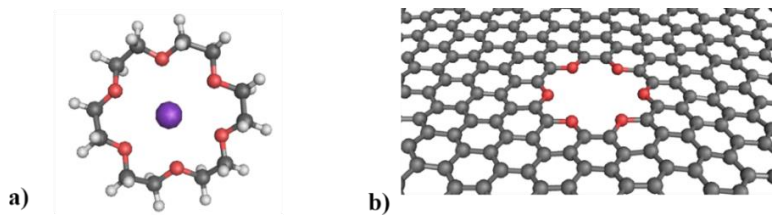

Figure 1. A stand-alone 18-crown-6 ether with an embedded potassium ion (a) and its graphene-embedded analog (b).

The hexagonal symmetry of graphene naturally lends itself to the possibility of embedding various types of crown-like pores in a graphene layer. Fabrication of such pores *via* vacancy oxidation was recently reported.[6] In particular, a hexagonally symmetric 18-crown-6 pore in graphene is shown in Fig. 1 (b). It can be viewed geometrically as a result of removing an entire carbon hexagon, followed by replacing the remaining edge carbons with oxygen atoms. From the chemical standpoint, introduction of an 18-crown-6 pore results in formation of additional defects (or healing of existing defects) in the graphene sample – see section S4 of the Supporting Information (SI). Such a pore is expected to exhibit significant binding preference for aqueous $K^+$ ions, as shown in the aqueous titration studies of stand-alone crown ethers[7] and from density functional theory (DFT) calculations performed in vacuum for the graphene-embedded 18-crown-6 structure analog.[6] Given the potential utility of ultra-narrow nanopores in atomically thin membranes for applications ranging from water filtration and sensing[8-9] to energy storage,[10-11] these membrane-embedded pores are expected to open various additional possibilities.

All liquid-phase studies to date have been focused on the cation-binding properties of stand-alone crown ethers. Currently, given a realistic possibility of fabricating atomically thin membranes with embedded crown ethers[6] and similar structures,[12-13] ion-binding and permeation properties of such membranes may prove promising for a wide range of applications. Here, with the use of molecular dynamics (MD) simulations, we investigate aqueous binding and transport properties of the 18-crown-6 pores embedded in monolayer graphene, immersed in aqueous solutions of NaCl and KCl of varying concentrations. Our results confirm zero anion permeability of the ether-porous membranes, as well as demonstrate semiconductor-like transport of $K^+$ ions, while transport of $Na^+$ ions follows the expected Michaelis-Menten saturation.[14] Our findings show potential promise of membrane-embedded crown ether pores for numerous applications, including water filtration and fundamental studies of ionic hydration at the deep nanoscale. Importantly, we demonstrate that simple ion-based logical operations can be performed using graphene-embedded crown pores.

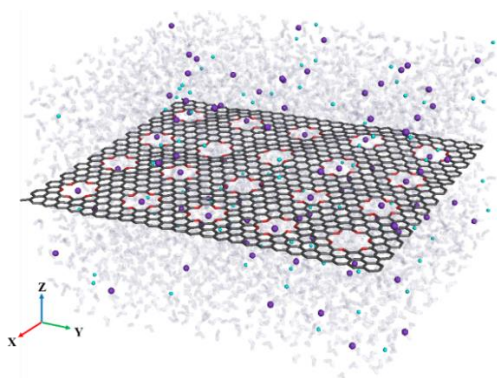

Figure 2. An example of the simulated system: graphene membrane with an embedded array of 20 18-crown-6 ethers, in an aqueous ionic bath. Dissolved ions are shown as solid spheres and water molecules are transparent for clarity. The system dimensions are 5.5 nm × 6.4 nm × 5.0 nm.

**Results and discussion**

The model systems investigated in this work are based on a 5.5 nm × 6.4 nm suspended graphene sheet with one or multiple embedded 18-crown-6 pores. The sheet was restrained along its perimeter and immersed in a rectangular periodic aqueous ionic bath (Fig. 2).

We begin by investigating aqueous ion dynamics in the presence of a room-temperature graphene membrane with nine embedded 18-crown-6 ether pores. Separate simulations of 0.15 M KCl and NaCl were performed for 100 ns. A representative initial state of the system is shown in Fig. 3 (a). In the case of KCl, all pores become populated by $K^+$ ions (Fig. 3 (b)) within the first 50 ns of the simulated time and such a configuration remains stable throughout the remainder of the simulation. In contrast, this behavior is not observed for NaCl, as expected from the 18-crown-6's binding preference for potassium cations in water.[7] We consider the energetics of the observed ion assembly and trapping further by combining umbrella sampling simulations and the weighted histogram analysis method (WHAM)[15] to extract the graphene-embedded crown ether's binding Gibbs free energy to aqueous $K^+$ and $Na^+$ ions.

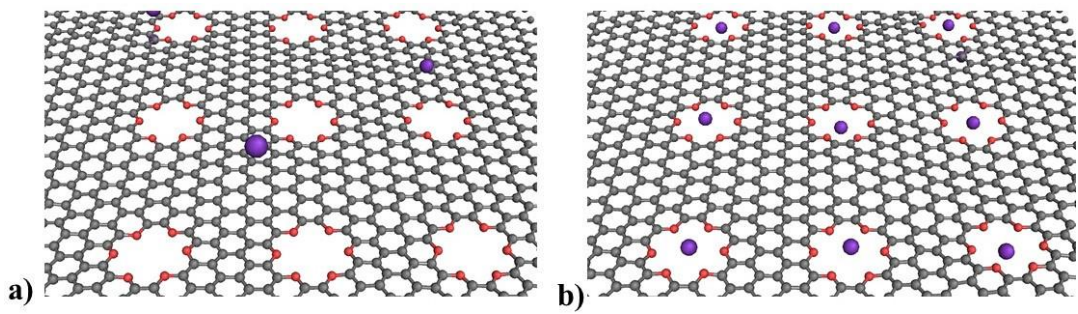

Figure 3. Initial (a) and final (b) state of a graphene membrane featuring an array of nine ether pores, as obtained in a room-temperature molecular dynamics simulation in aqueous 0.15 M KCl solution. No stable ion trapping was observed in an identical system in NaCl solution.

The results for the binding Gibbs free energy ΔG as a function of "reaction coordinate" (here, distance along the Z-axis from the membrane) are shown in Fig. 4 (a). For both cation species, we observe ΔG minima and maxima corresponding to ions' interaction with the pore, as well as the hydration-induced transfer barrier, essentially delineating the transition of the ion from bulk solution to the pore confinement. The Gibbs free energy minima corresponding to the equilibrium binding site are -14.4 kJ/mol and -3.8 kJ/mol for $K^+$ and $Na^+$, respectively – in reasonable agreement with the corresponding calorimetric data of -11.72 kJ/mol and -4.79 kJ/mol obtained for a stand-alone 18-crown-6 ether.[7] As expected, a significantly larger affinity for $K^+$ ions is observed, so that the probability of trapping $K^+$ ions is $\exp\left(\frac{\Delta G_{K+} - \Delta G_{Na+}}{k_b T}\right) \approx 25.5$ times higher than that of $Na^+$ ($k_b$ and $T = 300$ K are the Boltzmann constant and the system temperature, respectively). Above, $\Delta G_{Na+} = 12.4$ kJ/mol and $\Delta G_{K+} = 20.5$ kJ/mol are the "peak-to-peak" Gibbs energies, including the transfer barriers in Fig. 4 (a). The significant planarity-induced binding enhancement in vacuum relative to stand-alone ethers[6] is not observed for a membrane suspended in room-temperature solvent. Solvent screening is expected to account for most of the binding energy reduction in our simulations, as well as in calorimetric measurements of free aqueous crowns. In addition, the static density functional theory calculations[6] considered a planar graphene structure, and thus *flexural* relaxation of the pore region necessary for suspended graphene at finite temperature was excluded. At the same time, immediate trapping of $K^+$ ions and a low "bulk-to-pore" transfer barrier shown in Fig. 4 (a) are consistent with predictions in,[6] as graphene-embedded crowns indeed do not require significant radial stretching to trap an ion.

It is worth noting that the ΔG minimum for $Na^+$ in Fig. 4 (a) is located away from the plane of the ether-containing membrane. To investigate this observation further, we plotted the cation-water radial distribution functions *g(r)* for each ionic species inside and outside of the crown ether pore. The results are shown in Fig. 4 (b), where the distributions for $Na^+$ inside the pore and in bulk solution are quite similar, whereas for $K^+$ a considerable reduction in hydration is observed inside the pore. Given a generally more stable hydration shell around $Na^+$,[16-17] non-preferential binding and thus a "wobbly" host-guest interaction for the $Na^+$ ion results in slight deformation of its hydration shell while the ion resides in the ether pore. The presence of the deformed hydration shell around $Na^+$ ion then results in the corresponding $\Delta G$ minimum position away from the membrane plane. The observed contrast with the "tight fit" observed between $K^+$ and the 18-crown-6 ether is consistent with earlier results[18] and, together with the considerably larger absolute value of $\Delta G$ minimum for $K^+$, can be viewed as a clear manifestation of the host-guest ion recognition in the case of an ether embedded in a membrane.

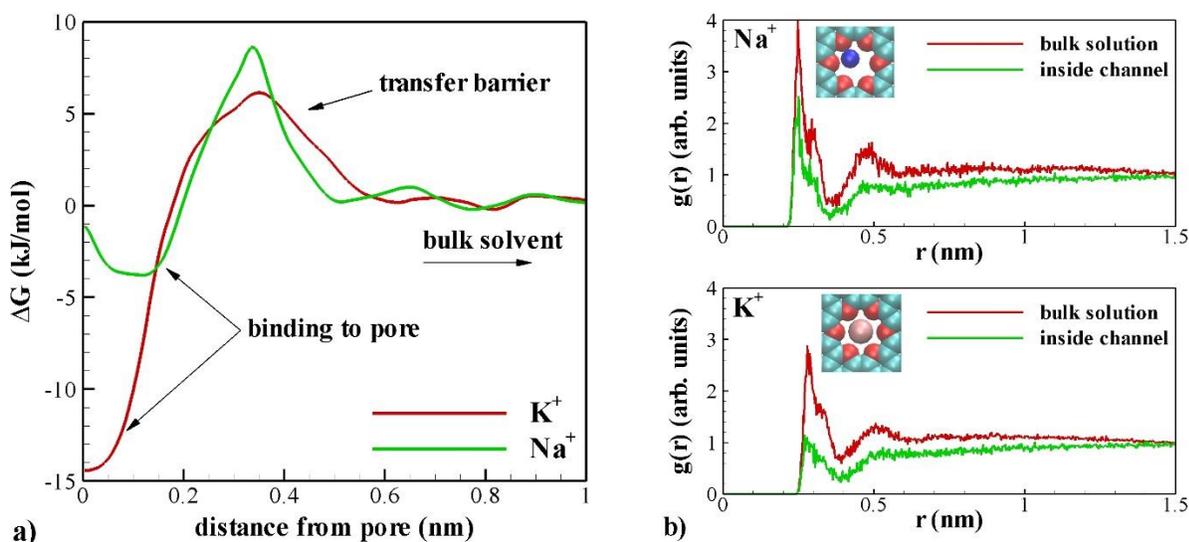

Figure 4. MD-simulated binding Gibbs free energy as a function of ion's distance from the center of the crown-like nanopore (a) and cation-water RDF curves for $Na^+$ and $K^+$ (b).

It is important to note that the highly selective $K^+$ trapping behavior observed here is beyond merely capturing ions from the solution. After effectively becoming a stable part of the membrane, potassium ions not only block the membrane for permeation, but also result in a "macroscopic" solution-exposed sheet charge, leading to a long-range electrostatic barrier in addition to those discussed above (Fig. 4). The effect of this sheet charge on the system energetics can be estimated. The magnitude of the effective electric field induced by a sheet charge density $\sigma$ is $E = \sigma/2\varepsilon\varepsilon_0$, where $\varepsilon$ and $\varepsilon_0$ are the effective dielectric constant of water near the membrane and the vacuum permittivity, respectively. With a total of nine K$^+$ ions embedded in a 5.5 nm × 6.4 nm membrane, the sheet charge density is 0.04 C/m$^2$ and thus with bulk $\varepsilon = 80$ we find $E \approx 30$ MV/m. For example, at a distance of 1 nm away from the membrane, such a field introduces a non-negligible potential difference of 30 mV relative to the membrane plane. In simulations of a membrane containing a total of 20 embedded crown ether pores (system shown in Fig. 2), similar K$^+$ trapping was observed, increasing the sheet-induced field estimate above by a factor of greater than two. Note that the presented calculations likely underestimate the effective local field, because $\varepsilon$ of a polar liquid is known to be spatially distributed near a charged solid surface and *reduced by as much as an order of magnitude* in the 0.2-0.3 nm wide interface region, compared to the bulk value of 80.[19] The implications of the sheet charge induced by the trapped ions also result in interesting possible applications, as discussed later.

Resulting from the field induced by the trapped charge, an electric double layer (EDL) is expected to form in an aqueous salt solution between the ion-containing membrane and mobile counterions. EDL formation in the case of KCl is confirmed in Fig. 5, where we plot the time-averaged ion densities (averaged per XY-slice, each parallel to the membrane) as functions of distance along the Z-axis for different ion species. In contrast, because the membrane occupancy by Na$^+$ ions is significantly lower, virtually no EDL formation is observed for NaCl, and overall considerably weaker charge polarization is observed throughout the interfacial region.

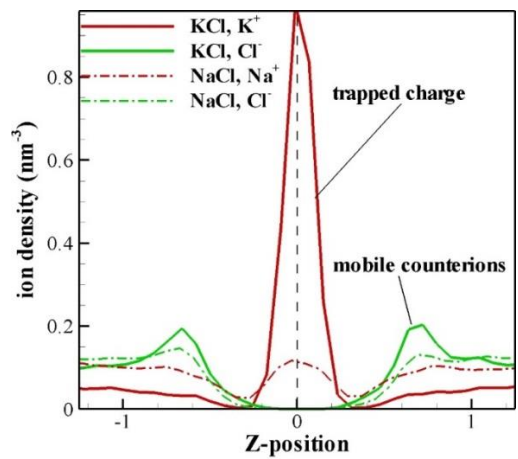

Figure 5. Ion density distributions in 0.15 M simulations of NaCl and KCl. The membrane is positioned at Z = 0.

EDL formation essentially replaces the short-range van der Waals interactions between ions and the individual unoccupied pores with longer-range interactions. Interestingly, the system presented here can be viewed as somewhat of an opposite of the porous graphene considered earlier, where *unoccupied* pores with carboxyl interiors exhibited no ion trapping behavior and carried pH-dependent electric charge.[20] It is also different from the protein ion channels embedded in relatively thick lipid bilayers and featuring buried charges responsible for their selectivity. We see that the effective ether pore occupancy (the average time an ion is trapped by the membrane) is responsible for the presence of the EDL at steady-state, and thus for the interactions in the system, as well as possible effects thereof on the ionic transport. As a result, one can expect significant differences in ionic permeation and conduction selectivity, depending on the ion species. Ion transport induced by externally applied transmembrane fields is considered next.

We performed a series of MD simulations of ionic transport *via* crown ether pores embedded in graphene membranes. Different NaCl and KCl concentrations (0.15 M, 0.5 M, and 1 M) were simulated. Each simulated current value was obtained from a 200 ns-long simulation and an external bias in the form of a constant electric field was applied perpendicularly to the membrane (along the Z-axis). The external field magnitude $E_{ext}$ varied in the range 0-100 MV/m in 10 MV/m increments. For simplicity of presentation, we report all currents as functions of the "transmembrane" potential drop across the simulated system height (along the Z axis), which in this case is $E_{ext} \times h$, where $h$ = 5 nm. The per-channel current at each value of transmembrane potential was calculated from the slope of a straight line fitted to the cumulative ion flux *via* membrane, rescaled by the number of channels (also, see section S1 of the SI). The results are shown in Fig. 6.

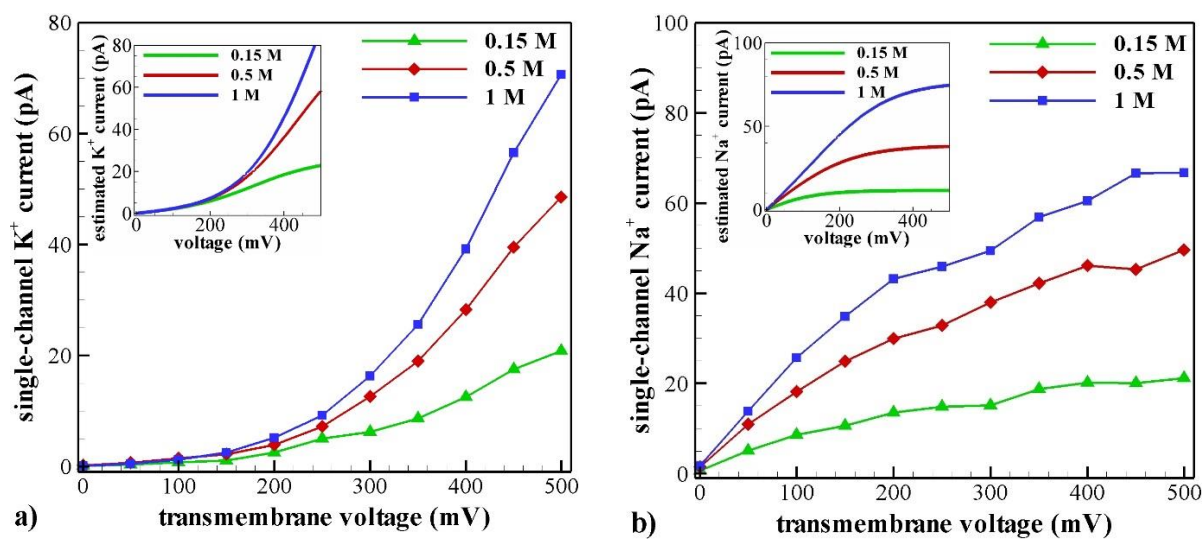

Figure 6. Simulated current-voltage curves for K$^+$ (a) and Na$^+$ (b); no anion transport was observed regardless of concentration, or the external bias magnitude, suggesting that 18-crown-6 pores are infinitely cation-selective. The insets show corresponding current estimates obtained from the theoretical model described in the text.

For KCl, virtually no current is observed in Fig. 6 (a) at voltages ≤ 150 mV and the membrane is in the "off" permeation mode *regardless of the concentration*, followed by a transition to the "on" mode at voltages ≥ 300 mV. Essentially, at low voltages the membrane remains blocked by the trapped ions, while the process of knocking out the trapped ions by the incoming mobile ions is likely suppressed by the accompanying presence of the EDL. As the transmembrane voltage increases, the probability of thermally induced dissociation of the trapped ions increases, while knock-on events also become more prominent, once again compounded by the weakening EDL (the effective occupancy decreases, thus reducing $\sigma$). The results for NaCl in Fig 6 (b) are different qualitatively, because no ion trapping occurs and the additional effects of the EDL are less pronounced – we revisit this point in greater detail later when discussing the data shown in the Fig. 6 insets. Given the simulated data in Fig. 6, 18-crown-6 pores are sodium-selective in terms of permeation, as shown in Fig. 7. For low voltages, the selectivity ratio varies from 20 to 35, eventually approaching unity at high bias values, as the membrane occupancy by $K^+$ decreases (see inset of Fig. 7) and the conduction regimes become identical for both ionic species.

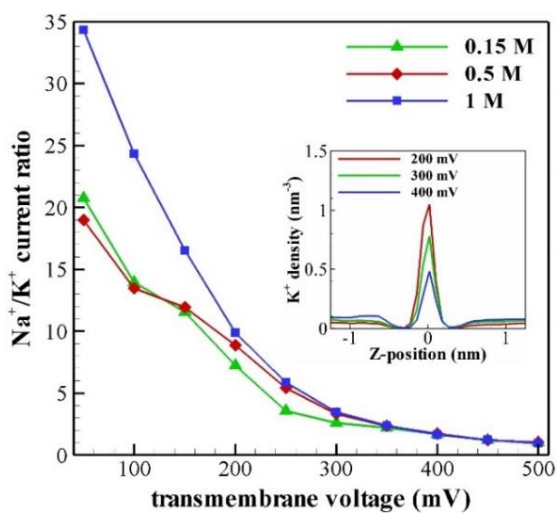

Figure 7. $Na^+/K^+$ current selectivity for various salt concentrations, as obtained from the data presented in Fig. 6. The inset shows $K^+$ ion distributions at different transmembrane voltages for a 0.15 M KCl simulation.

Our ionic permeation findings are in contrast with those obtained for graphene-embedded carboxyl pores,[20] where qualitatively similar trends were observed for $Na^+$ and $K^+$. The current-voltage trends observed here, however, can be qualitatively explained with the use of existing theoretical treatment of reaction kinetics. Although most analytical models lack a detailed treatment of the interatomic interactions, hydration effects, mechanisms of mobile ions knocking out trapped ions, or the effects of the EDL, qualitative insight can still be obtained for illustration. Consider the association-dissociation model described in detail elsewhere.[21] For the 18-crown-6 pores, strictly one ion permeates at a time, which allows for defining distinct binary occupancy states of the channel. Assuming ionic concentration symmetry above and below the membrane, the resulting current is

$$I = \frac{qk_d c \sinh\left(\frac{qV}{2k_b T}\right)}{K_d \cosh\left(\frac{qV}{2k_b T}\right)+c}, \quad (1)$$

where $q$ is the cation charge, $k_d$ is the ion-pore dissociation rate, $K_d = k_d/k_a$ is an effective constant determining the association rate $k_a$ for each ionic species, and $c$ and $V$ are the salt concentration and transmembrane voltage, respectively. Mathematically, the expression above is of the Michaelis-Menten type,[14] with saturation occurring when the association term in the denominator begins to approach the voltage-dependent dissociation term in the numerator at a given concentration and voltage.

With $k_d = 1.2 \times 10^7$ s$^{-1}$, $K_d = 0.011$ M and $k_d = 1.8 \times 10^8$ s$^{-1}$, $K_d = 0.37$ M selected as rough fitting parameters for K$^+$ and Na$^+$, respectively, theoretical estimates for each ion species are provided in the corresponding insets of Fig. 6 (a) and (b). Because our aim here is basic illustration, we deliberately avoid plotting the theoretical estimates alongside the simulated data, or fine-tuning the model parameters to quantitatively fit the simulated current-voltage curves. Theoretical curves reveal that the qualitative difference between K$^+$ and Na$^+$ permeation trends is due to the large difference in the ion-pore dissociation rates. For a channel that allows permeation of only one ion at a time, the current is essentially $q/\tau$, where $\tau$ is the effective average ion transition time. The latter can be defined as $\tau = \tau_d + \tau_a$, where $\tau_d$ is the average time an ion spends trapped in the pore (dissociation time) and $\tau_a$ is the salt concentration dependent effective time it takes for an ion to enter the pore from bulk solution (association time). For K$^+$ at low voltages, $\tau_d \gg \tau_a$ and thus, regardless of $\tau_a$ (and of the ionic concentration) $\tau \approx \tau_d$, confirmed in Fig. 6 (a). For Na$^+$, $\tau_d$ is considerably shorter (at zero bias, $\tau_{d,K+}/\tau_{d,Na+} = k_{d,Na+}/k_{d,K+} = 15$, of the same order as ~25.5 calculated earlier from the Gibbs energies), allowing considerable permeation at small voltages and further decreasing upon increasing bias. At high bias, $\tau_d$ eventually becomes smaller than $\tau_a$, which leads to association-limited saturation.

The distinguishable K$^+$ conduction regimes discussed above and especially the dependence of membrane effective ion occupancy on the transmembrane voltage (see inset of Fig. 7) immediately suggest an intriguing possibility of performing some elementary ion-based logical operations. Direct electrical measurement of the membrane potential is then effectively a read operation. As follows from earlier discussions, the time-averaged trapping occupancy (or charge) is estimated to be $\frac{q_0}{1+\frac{\tau_a}{\tau_d}}$, where $q_0 = +N \times e$ is the maximum trapped charge in a membrane with $N$ crown pores. At sufficiently low KCl concentration, such that $\tau_a \gg \tau_d$ at large transmembrane bias, the highly conductive ("on") regime coincides with low occupancy, and vice versa. The effect of trapped charge variation on the electrical potential in the region occupied by the membrane was simulated directly using MD at various transmembrane potentials. The time-averaged results are shown in Fig. 8 (a). As simulated, the membrane potential indeed decreases by more than 100 mV with increasing transmembrane bias; $|dV_{out}/dV_{in}|$ is shown to be lower at the higher salt concentration due to overall shorter association times $\tau_a$, as expected. Note that the trapping-induced maximum potential at low transmembrane bias (proportional to the number of pores) can be increased with a higher pore count (see section S2 of the SI). In the context of Figs. 6 (a) and 8 (a), consider a low transmembrane voltage $V_{in}$ (< 200 mV), denoted "0," applied to a cell with appropriately selected KCl concentration, as shown in the top pane of Fig. 8 (b). In this case, the cell is nearly non-conductive ("off") and the membrane is fully occupied by the trapped ions. Therefore, the potential $V_{out}$ measured directly at the membrane is relatively high, denoted as "1." Conversely, when high $V_{in}$ (> 300 mV), denoted "1," is applied, the cell is highly conductive ("on"), fewer ions are trapped in the membrane, and thus a low ("0") state of $V_{out}$ is measured. With appropriately set "0/1" thresholds, the input-output relationship here can be viewed as the Boolean NOT operation. As a straightforward extension, exclusive OR (XOR) operation is possible with two independently driven cells shown in the bottom pane of Fig. 8 (b). When measuring the absolute value of differential $V_{out}$ between the two membrane states, one indeed observes state "1" only when *either of the two cells* is highly conductive, corresponding to a XOR operation.

The effective state switching speeds can be estimated. During the "on-off" transition, $K^+$ ions begin to populate the pores, limited by the ionic diffusivity and thus the low-bias association time $\tau_{a,lb}$. Switching in the opposite direction is limited by the high-bias dissociation time $\tau_{d,hb}$. The order of magnitude for $\tau_{a,lb}$ and $\tau_{d,hb}$ can be estimated from the permeation data in Fig. 6 as $(q/I)$ at corresponding bias strengths, where $q$ is the cation charge and $I$ is the single-channel current (also, see supplementary movie of $K^+$ ions trapping in an initially empty membrane). For the pore spacing presented here, at 0.5 M KCl the characteristic time limit $max(\tau_{a,lb}, \tau_{d,hb})$ is of order of a few nanoseconds, corresponding to hundreds of MHz.

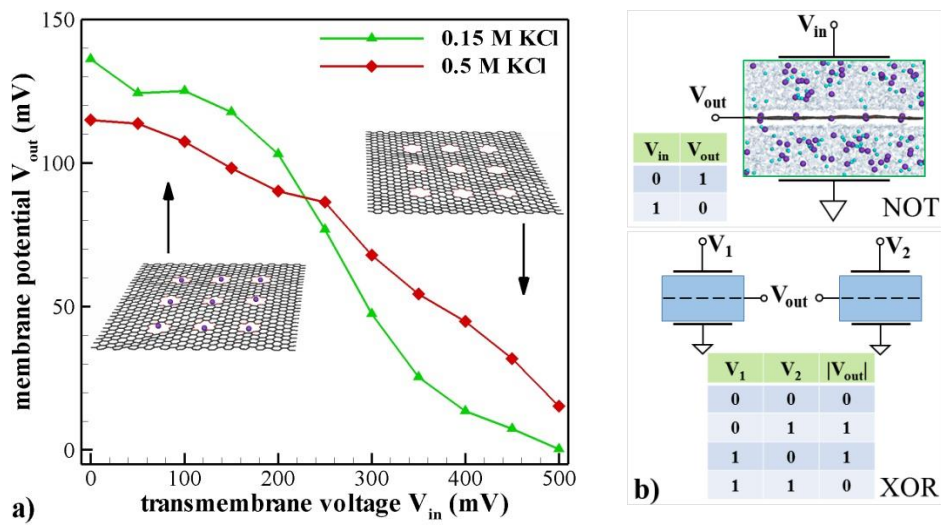

Figure 8. Membrane potential as a function of transmembrane voltage, simulated at 0.15 M and 0.5 M KCl (a), and simple NOT and exclusive OR (XOR) logical operations obtained from measuring the effective membrane potential in the "on" and "off" permeation modes (b). For clarity, a baseline potential of ≈ 550 mV, arising from water ordering at the graphene-water interface [22], was subtracted from all simulated Y-axis values in (a) to yield potential close to zero at the highest transmembrane voltage, corresponding to the lowest ion occupancy.

The ion-based logic examples presented above are attractive due to their conceptual simplicity, compared with the devices proposed earlier,[23-24] especially given that the distinction between the "on" and "off" permeation regimes is achievable with voltage magnitudes induced by salt concentration gradients often found in biological systems. More sophisticated cascaded logic with ionic diodes and transistors may then be high interest. For instance, in multilayer systems featuring membranes with different numbers of embedded crown ether pores, the resulting asymmetry (including electrostatic asymmetry) suggests the possibility of diode-like current rectification, similar to nanofluidic diodes described earlier.[25-28] Here, we explore the possibility of electrostatically gating the cell shown in Fig. 8, essentially resulting in a simple ionic transistor.

In the case of ultra-narrow pores in graphene (as well as in other conductive atomically thin materials), immediate exposure of the membrane to the surrounding ionic solution suggests an opportunity for control of the EDL and thus for gated ionic flow. The control voltage is applied directly to the membrane, as described schematically in Fig. 9 (a). In contrast with buried-gate ionic transistors featuring field-induced narrowing of the relatively wide conducting channels [29-30], the effective "supply" of ions to the pores (*i.e.* the effective association rate in Eq. (1)) is subject to control here.

To assess tuneability of the ionic current, we mimicked a gate voltage applied directly to the graphene membrane. Because it is impossible to set specific voltages in simulations involving periodic boundaries and Ewald summation for solving Poisson's equation, we did so by simulating delocalized excess charge externally injected into the membrane in the form of a small nonzero charge of every "bulk" carbon atom. However, we show below that it is possible to analytically estimate the effective gate voltage associated with this excess charge. The resulting system was neutralized by modifying the number of Cl$^-$ ions accordingly. The number of dissolved cations was left unchanged for direct comparison of the current with the floating gate case (corresponding data from Fig. 6). The effective "excess" gate voltage variation (in addition to that arising from the charge induced by trapped ions, which, as discussed earlier, depends on the transmembrane voltage) relative to "drain," can be estimated analytically: $V_g = \frac{\Delta Q}{2A\varepsilon\varepsilon_0} \times \frac{h}{2}$, where $\Delta Q$ and $A$ are the total externally induced charge and the area of the membrane, respectively. For a relatively low $|\Delta Q| = 10e$ and 1136 "bulk" carbons in the simulated membrane, the per-carbon charge is ~$8.803 \times 10^{-3}$e. The corresponding "excess" gate voltage $V_g$ is then estimated to be 80 mV ≈ $3k_bT$. The current-voltage curves for 0.5 M KCl and $\Delta Q = \pm 10e$ were simulated in two series of independent 200 ns long simulations. The results, along with the $\Delta Q = 0$ case (0.5 M KCl data from Fig. 6 (a)), are shown in Fig. 9 (b).

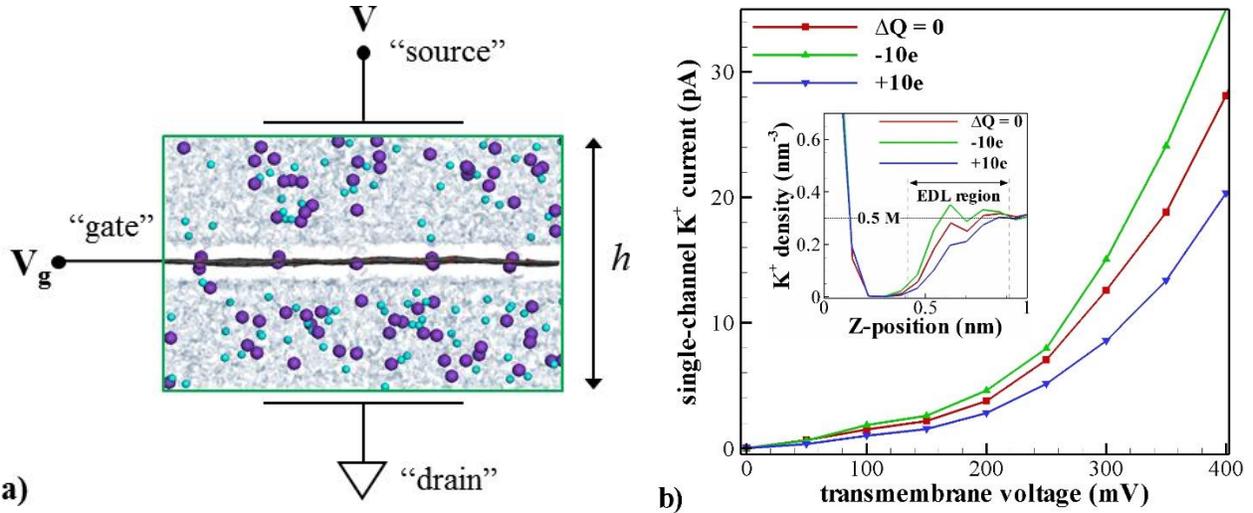

Figure 9. Ionic transistor schematic (a) and current-voltage curves (0.5 M KCl) for various values of $\Delta Q$ (b). The inset in (b) shows $K^+$ density distributions at various $\Delta Q$ and zero transmembrane voltage.

As shown, current modulation is significant even with a gate voltage variation of the order of a few $k_b T$. For example, at the transmembrane voltage of 300 mV, the estimated effective *per-channel* transconductance is >70 pS, suggesting that with an appropriately selected transistor load and an accurate "on/off" output voltage threshold, sensitive switching may be possible. As shown in the inset of Fig. 9 (b), the mobile $K^+$ population adjacent to the membrane, central to the gating effect on current in this system, is indeed modulated. Qualitatively, it may be viewed as replacing the value of $c$ in Eq. (1) with $c \times \exp\left(-\frac{qV_g}{k_b T}\right)$. It can be shown that, given Eq. (1), the transconductance $g = \frac{dI}{dV_g}$ exhibits resonant behavior with respect to $c$ and thus can be further optimized in terms of salt concentration for a selected transmembrane voltage (see section S3 of the SI). Overall, the observed current sensitivity to external gating is likely sufficient for implementing cascaded ion-based logic in the form of ionic circuit elements supplying control signals, as well as those subject to control.

**Conclusions**

We performed extensive MD simulations to investigate ion trapping and permeation in graphene-embedded 18-crown-6 pore arrays immersed in aqueous KCl and NaCl. In agreement with the experimental data for stand-alone 18-crown-6 ethers in water, we estimated the binding Gibbs energies at -14.4 kJ/mol and -3.8 kJ/mol for $K^+$ and $Na^+$, respectively. Spontaneous stable trapping of $K^+$ ions in the crown pores was shown at room temperature. Externally induced permeation across membranes featuring embedded crown ether pores reveals qualitatively different current-voltage characteristics for KCl and NaCl. For KCl, "on" and "off" conduction regimes exist due to the presence of an explicit permeation "gap" (in the form of the ion-pore binding energy), accompanied by the presence of EDL, arising from the trapped ionic charge, especially at low ($\leq 150$ mV) transmembrane voltages. For NaCl, no cation trapping occurs and Michaelis-Menten-type current-voltage behavior is observed. Given the $K^+$-selective trapping, permeation is shown to be highly $Na^+$-selective, with $Na^+/K^+$ selectivity ratio decreasing as the transmembrane voltage increases. For both NaCl and KCl, permeation is demonstrated to be infinitely cation-selective, regardless of salt concentration.

Interestingly, we demonstrated utilization of membrane charging by trapped $K^+$ ions for ion-based logic, and simple examples of NOT and XOR gates were presented. Together with a sensitive ionic transistor, also shown in this work, cascaded ion-based logic may be possible. Our results therefore suggest a host of potential applications. Stable ion trapping, chemically tunable to target various ion types, may be utilized for energy storage, highly selective gas- and liquid-phase ion sensing, as well as for deionization against specific ion species. Ether pores embedded in suspended graphene may also be used to study ionic solvation at a single ion resolution. Voltage-tunable trapping of ions in membrane-embedded crowns may hold promise for storing information, while sophisticated ion-based logic from combining conceptually simple state-switching ionic cells and ionic transistors may be achieved. Potential applications are of course not limited to graphene, and membrane-embedded crown-like pores in other atomically thin materials may be of even greater interest in terms of the structure-function relationship. Transition metal dichalcogenides and boron nitride are among the candidates for applications in liquid and gas phase, as a wide range of membrane pore structures and hydrophobicity levels is expected in these materials.

**Methods**

The intramolecular model for graphene was based on a harmonic potential informed by the optimized bond-order potential,[31] as utilized earlier.[32-34] All intermolecular interactions in the system were based on the well-established OPLS-AA forcefield,[35-36] including the partial atomic charges (see section S4 of the SI for details) and the Lennard-Jones parameters, responsible for the interactions between ions and the crown ether pores. The TIP4P model[37-38] was used to represent water molecules. All MD simulations were performed using GROMACS v. 2016.4 software[39-40] with GPU acceleration. Prior to production runs, all systems were pre-relaxed in NPT simulations at $T = 300$ K and $p = 0.1$ MPa with a time step of 1 fs. All production simulations were performed in an NVT ensemble with a time step of 2 fs for hundreds of nanoseconds, as specified in the text.

**Supporting Information**

Supplementary details of the simulated system, methods, and additional results and discussion (PDF).

Animation of MD-simulated atomic trajectories depicting 50 nanoseconds of room-temperature aqueous ion dynamics (0.15 M KCl; water molecules omitted from visualization for clarity) in the presence of a locally suspended graphene membrane with a total of nine 18-crown-6 ethers embedded. $K^+$ and $Cl^-$ ions colored white and blue, respectively (AVI).

**Acknowledgment**

We thank A. Fang, J. A. Lemkul, B. I. Yakobson, and A. N. Chiaramonti for useful discussions.

Authors gratefully acknowledge support from the Materials Genome Initiative. This work is a contribution of the National Institute of Standards and Technology, an agency of the US government. Not subject to copyright in the USA. Trade names are provided only to specify procedures adequately and do not imply endorsement by the National Institute of Standards and Technology. Similar products by other manufacturers may be found to work as well or better.

**Supporting information**

**Aqueous Ion Trapping and Transport in Graphene-embedded 18-crown-6 Ether Pores**

Alex Smolyanitsky, Eugene Paulechka, Kenneth Kroenlein

*S1. Ionic fluxes*

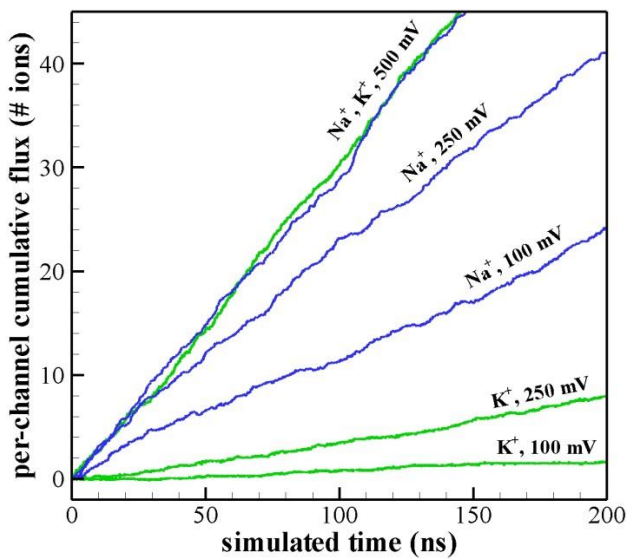

Figure S1. Typical cumulative fluxes obtained for different transmembrane voltages (0.5 M *KCl* and *NaCl*). Zero Cl$^-$ fluxes were obtained in all simulations.

*S2. Permeation and membrane potential in a 20-pore membrane*

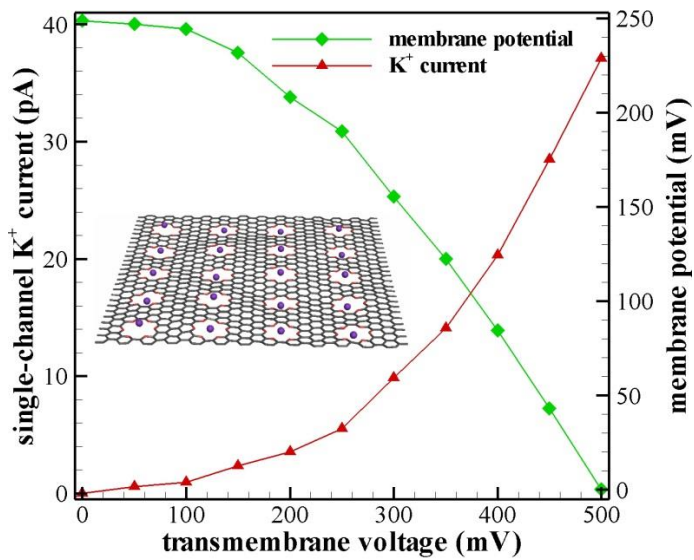

Figure S2. Simulated current and membrane potential as functions of the voltage across a 20-pore membrane (a fully occupied state is shown in the inset) for 0.5 M *KCl*. Each point was obtained from a 200 ns long room-temperature MD simulation, as described in the main text.

S1

## S3. Transistor sensitivity tuning

We rewrite Eq. (1) from the main text and replace the salt bulk concentration $c$ with $c \exp\left(-\frac{qV_g}{k_bT}\right)$:

$$I = \frac{qk_d c \exp\left(-\frac{qV_g}{k_bT}\right)\sinh\left(\frac{qV}{2k_bT}\right)}{K_d \cosh\left(\frac{qV}{2k_bT}\right) + c \exp\left(-\frac{qV_g}{k_bT}\right)} \quad (S1)$$

The differential per-pore transconductance is

$$g(c) = \frac{dI}{dV_g} = -\frac{q^2 k_d K_d}{k_b T} f_1(V)\, c \left(K_d f_2(V) + c \exp\left(-\frac{qV_g}{k_bT}\right)\right)^{-2}, \quad (S2)$$

where $f_1(V) = \sinh\left(\frac{qV}{2k_bT}\right)\cosh\left(\frac{qV}{2k_bT}\right)$ and $f_2(V) = \cosh\left(\frac{qV}{2k_bT}\right)$. The absolute value of $g$ is maximized at $K_d f_2 = c \exp\left(-\frac{qV_g}{k_bT}\right)$, yielding

$$c_0 = K_d \exp\left(\frac{qV_g}{k_bT}\right)\cosh\left(\frac{qV}{2k_bT}\right). \quad (S3)$$

With $k_d$ and $K_d$ used in the main text, we can evaluate Eqs. (S2) and (S3) for *KCl*. At an operating point of $V = 200$ mV and the current modulated around $V_g = 0$, we obtain $c_0 = 0.26$ M and $g(c_0) \approx -0.44$ nS. At a higher operating point of $V = 300$ mV, $c_0 = 1.78$ M and a considerably larger optimal transconductance is obtained: $g(c_0) \approx -3.0$ nS. Presented calculations are only accurate at the order-of-magnitude level, but clearly demonstrate that transistor sensitivity exhibits resonant properties with respect to salt concentration, and is realistically tunable for a selected operating transmembrane voltage $V$. *KCl* concentration yielding maximal absolute value of transconductance as a function of $V$ is shown in Fig. S2. The inset demonstrates resonant behavior of $|g|$ with respect to $c$ at $V = 200$ mV.

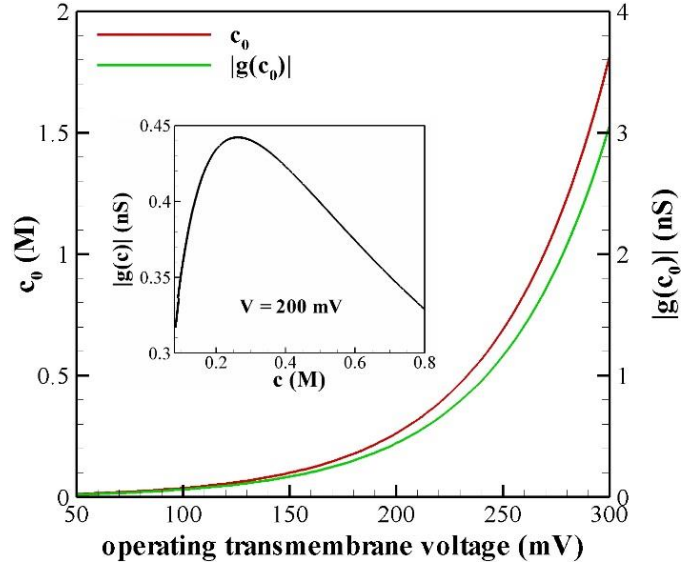

Figure S3. Optimal *KCl* concentration $c_0$ as a function of selected transmembrane voltage operating point around $V_g = 0$. The inset shows absolute value of transconductance $|g|$ as a function of bulk salt concentration for $V = 200$ mV with a maximum at $c_0 = 0.26$ M.

## S4. Structure and atomic charges in graphene-embedded 18-crown-6 pores

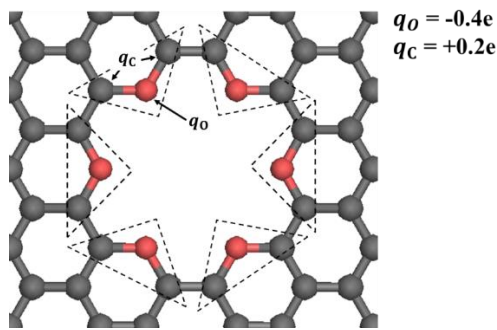

Figure S4. Atomic charges, according to the OPLS-AA forcefield; the circular charge distribution was confirmed by quantum-mechanical calculations. Each resulting dipole is outlined by a dashed triangle and the charges of each atomic species are equal between dipoles.

A direct replacement of $sp^2$ carbons with oxygens in an infinite pristine graphene sheet will necessarily introduce defects like uncompensated charge, or require extra hydrogenation to obtain system neutrality. However, in a finite sheet interfaced with a conducting substrate, however large, this situation can be avoided (see, for example, Fig. S5 with hydrogen-passivated edges).

A realistic graphene sample is expected to have naturally occurring defects, including C-H, C-O, or C=O bonds, which, along with the presence of mobile charge in graphene, can compensate for the defects from introducing the pore structure, thus maintaining electrical neutrality of the overall system. The correctness of this assumption is supported by Figure 2 of Ref. 6 in the main text, where most of the experimentally obtained oxygen-containing rings are shown to contain six atoms. As a result, a simplified model presented here should be able to describe the main features of the crown-porous membrane in terms of its interactions with aqueous ions. In addition, such a simplification is not expected to produce a noticeable effect on the parameters used in the MD simulations. At the same time, more rigorous calculations (for example, using density functional theory) require explicit consideration of the defects. All charges shown in Fig. S4, along with the planarity of the pore region, were confirmed by independent quantum-mechanical calculations (CHELPG scheme[1] at the *HF/6-31+G(d)* theory level, using Gaussian 09 software[2]) performed on a structure with defects similar to those shown in Fig. S5.

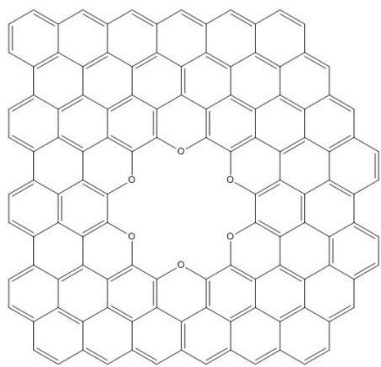

Figure S5. Bond distribution in a finite graphene sample with hydrogen-terminated edges and an embedded 18-crown-6 pore.